\documentclass{ws-ijmpa}
\usepackage[super,compress]{cite}
\usepackage{graphicx}
\begin{document}
\markboth{ D. Kirilova, V. Chizhov }{Chiral Tensor Particles in the
Early Universe}

%
%

\title{Chiral Tensor Particles in the Early Universe - Present Status}

\author{D. P. Kirilova}

\address{Institute of Astronomy and National Astronomical Observatory, BAS,
1784 Sofia, Bulgaria, dani@astro.bas.bg}

\author{V. M. Chizhov}

\address{Faculty of Mathematics, Sofia University,
1164 Sofia, Bulgaria}

\maketitle

\begin{history}
\received{Day Month Year}
\revised{Day Month Year}
\end{history}

\begin{abstract}
In this work an update of the cosmological role and place of the
chiral tensor particles in the Universe history is provided. We
discuss an extended model with chiral tensor particles.
     The influence of these particles on the early Universe evolution is studied.
      Namely, the increase of the
     Universe expansion rate caused by the additional particles in this extended model is calculated, their
     characteristic interactions with the particles of the hot Universe plasma are studied and the corresponding times of their
     creation, scattering, annihilation and decay are estimated for accepted
values of their masses and couplings, based on the recent
experimental constraints. The period of abundant presence of these
particles in the Universe evolution is determined.

\end{abstract}

\ccode{PACS numbers: 98, 14}


\section{Introduction}

 The chiral tensor (ChT) particles were first predicted
from theoretical considerations~\cite{MPLA} as an extension of
Standard Model. Namely, these new type of spin-1 particles complete
the set of Yukawa interactions and allow to realize all possible
irreducible representations of the Lorentz group. These chiral
tensor particles belong to the fundamental representation of
$SU(2){_L}$ group of the Standard Model. The ChT particles are boson
particles and they were predicted to be the carriers of new
interaction, however in contrast to the
   gauge bosons, they have only chiral interactions with the known fermions, through tensor
   anomalous coupling. For more detail
see the reviews~\cite{EChAYa,EChAYaa}.

    The chiral tensor particles may be produced and
detected at powerful high energy colliders. ChT particles have
unique properties, which will help to disentangle
    them from other widely discussed hypothetical particles at hadron
    colliders~\cite{LHC,mih2009}.
    Besides, the inclusion of the ChT
particles helps to solve the hierarchy problem~\cite{dvali}.

    At present
    the search of the chiral bosons is conducted by the international collaboration ATLAS
    at the Large Hadron Collider at CERN. After the first run of the
LHC experimental constraints on their masses were
obtained~\cite{ATLAS14,ATLAS14a}.

The cosmological influence of ChT particles has been first
considered by Kirilova et al, 1995 ~\cite{todor} and in later
publications ~\cite{AATr},~\cite{blois},~\cite{IJMPA} . It was found
that the new ChT particles contribute to the matter tensor in the
right-hand side of the Einstein--Hilbert equation, increasing the
Universe density and changing the dynamical evolution of the
Universe. Besides, possible direct interactions of ChT particles
were proposed with the particles present at the early stage of the
Universe evolution. The dynamical effect and the interactions of ChT
particles with the constituents of the early Universe plasma were
explored. A cosmological constraint on the strength of their
interaction $G_T<10^{-2}G_F$ was obtained~\cite{blois} on the basis
of BBN considerations~\cite{dol}. Hence, the ChT particles
interactions are expected to be centi-weak, which is in accord with
the theoretical and experimental findings.

Here we reconsider the
    processes involving ChT particles, namely their dynamical
effect and their interactions in the early Universe plasma using the
latest experimental constraints on their characteristics, that have
been obtained at ATLAS.

The next section describes the characteristics of the chiral tensor
particles. The third section presents an update of the cosmological
influence of ChT particles, namely an update of  their creation,
decay, annihilation and scattering processes and of the
characteristic scale of their typical processes and their dynamical
effect. The last section presents the conclusions.

\section{Chiral Tensor Particles Characteristics}

ChT particles are described by an {\em antisymmetric} tensor fields
of a rank two. They have a chiral charge and change the fermion
chirality. They have an anomalous (Pauli) interaction with matter.

 The ChT
particles are introduced as doublets $\left(T^+_{\mu\nu}
T^0_{\mu\nu}\right)$, like the Higgs particles. The possibilities
for new chiral anomalies are avoided by introducing an additional
doublet $\left(U^0_{\mu\nu} U^-_{\mu\nu}\right)$ with an opposite
chiral- and hypercharge. Correspondingly  also the Higgs sector is
increased: it becomes $\left(H^+_1 H^0_1\right)$, $\left(H^0_2
H^-_2\right)$.

   As in the Standard Model a Higgs-like mechanism is used to provide mass to the massless tensor particles,
  which have just longitudinal degrees
of freedom. The role of the Higgs field is played by a triplet,
denoted by $C_{\mu}$, and singlet gauge vector particles or by four
$SU(2)_L$ singlets (depending on the chiralities of the ChT
particles), $P^i_{\mu}$ ($i=1,...,5$), which allow them to acquire
transverse physical degrees of freedom.

 To avoid flavor violation in the neutral sector due to the
doubling of doublets it is assumed that the doublets $H_1$ and
$T_{\mu\nu}$ interact only with down-type fermions, while the
doublets $H_2$ and $U_{\mu\nu}$ -- with up-type ones~\cite{2higgs}.

\subsection{Chiral tensor particles degrees of freedom}

In the extended model with ChT particles their effective number of
the degrees of freedom changes. We have recalculated the degrees of
freedom, using consistent quantization of ChT fields with pyramid of
ghosts, as proposed by Chizhov and Avdeev~\cite{Avdeev}.  The
presence of the two additional tensor doublets, the triplet and
singlets gauge vector particles and the extra Higgs doublet
increases the total effective number of the degrees of freedom by
$$g_{ChT}=g_T+g_U+g_C+g_P+g_H=4+4+6+10+4=28.$$
Hence, the total number of the degrees of freedom, while the
additional particles are relativistic, is:
\begin{equation}
 g_{*}=g_{\rm SM}+g_{ChT}=106.75+28=134.75.
\end{equation}

\subsection{Chiral tensor particles masses}

 The presence of the vacuum expectation values of the two
different Higgs doublets leads to different masses for the tensor
particles interacting with up- and down-type fermions. Present
experimental constraints on the masses of the tensor particles
interacting with down type fermions at 95$\%$ CL are: $M_{T^0}>2.85$
TeV and $M_T^+>3.21$ TeV ~\cite{ATLAS14,ATLAS14a}, which are
considerably higher than assumed in previous publications.
Constraints on the masses of tensor particles interacting with
up-type fermions are not available yet.

\section{Cosmological Effects of the Chiral Tensor Particles}

 We reconsider the following cosmologically effects of ChT particles: their
 influence on Universe expansion rate due to the increase of the energy
density and ChT particles direct interactions with fermions.

\subsection{ChT particles influence on the Universe expansion}

Due to the additional particles in the extended model with ChT
particles, the energy density of the Universe is increased in
comparison with the Standard Cosmological Model case:
\begin{equation}
\rho=\rho_{SCM}+\rho_{ChT}.
\end{equation}
where the ChT particles contribution, while they are relativistic,
is: $\rho_{ChT}=\frac{\pi^2}{30}\:g_{ChT}\: T^4$, $T$ is the photons
temperature. Hence, the expansion rate of the Universe is increased:
\begin{equation}
H=\sqrt{8\pi^3 G_N g_*/90}\:T^2
\end{equation}
$g_*$ is the total number of the effective degrees of freedom from
(1). The temperature-time dependence is changed, correspondingly,
$t\sim1/(\sqrt{g_*}\:T^2)$.

\subsection{ChT particles interactions in the early
Universe}

 The tensor particles are supposed to
have interactions with the fermions.

At early epoch while ChT particles were relativistic
 their cross
sections decrease with energy increase, $\sigma\sim E^{-2}$, and
 hence ChT particles have been frozen. With the decrease of the
 Universe temperature in the course of the expansion the mean energy of particles also decrease and ChT
 interactions unfreeze when their characteristic interaction rates
$\Gamma_{int}\sim \sigma n$ become greater than the expansion rate
$H(T)$. The temperature of unfreezing $T_{eff}$ of an interaction
$i\rightarrow f$ is estimated from:
\begin{equation}
\sigma_{if}(T_{eff})n(T_{eff})=H(T_{eff})
\end{equation}
  The
corresponding cosmic time in seconds is:
\begin{equation}
t_{eff}\approx 2.42/\sqrt{g_*}\:T^2_{eff},
\end{equation}
where  $T_{eff}$ is in MeV.

Using the recent experimental constraints on the characteristics of
ChT particles, we provide an update of their creation, scattering,
annihilation and decay processes and estimate the characteristic
temperatures and cosmic times of these processes in the early
Universe.

{\it ChT particles creation from fermion-antifermion collisions}

\begin{figure}
  \centering
  \includegraphics[width=.25\textwidth]{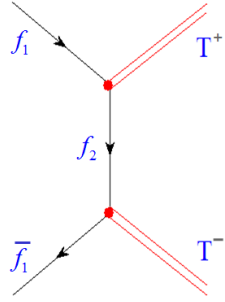}
  \caption{ChT particles creation from fermion-antifermion collisions}
\end{figure}
The creation of pairs of longitudinal
 tensor
particles from fermion-antifermion collisions has a cross-section:
\begin{equation}\label{cr}
    \sigma_c\approx\frac{g_T^4 \ln(T/v)}{4^5 \pi T^2}
\end{equation}
where $g_T$ is the ChT particles coupling and the Higgs vacuum
expectation is $v\approx 246$~GeV.

 We have found that the tensor
particle creation processes unfreeze when the Universe temperature
falls below $T_c$, where
\begin{equation} T_c\approx
1.83\times 10^{17}{~\rm GeV}
\end{equation}
This temperature is slightly higher (by an order of magnitude) than
previously estimated.

Thus, the ChT particles are created in the period when the Universe
temperature falls from $T_c$ till $T \sim 2M_T$.
 The
corresponding time of the unfreezing of ChT particles creation is:
$t_c\approx 6\times 10^{-42}$~s. It is earlier than previously
estimated.

 {\it
Fermions scattering on ChT particles}

\begin{figure}
  \centering
  \includegraphics[width=.35\textwidth]{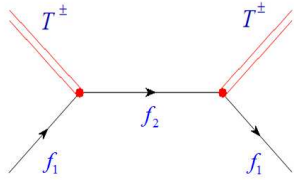}
  \caption{Fermions scattering on ChT particles}
\end{figure}
The cross-section of fermions scattering on ChT particles is given
by:
\begin{equation}
    \sigma_s\approx\frac{\pi g_T^2}{3 \times 4^5 \pi T^2}
\end{equation}
This process unfreezes at
\begin{equation}
T<T_s\approx 1.89\times 10^{15}~{\rm GeV}
\end{equation}.

$T_s$ is slightly changed (higher by an order of magnitude) compared
to previous estimation.
 The corresponding cosmic time period is $t>t_s\approx 5.86\times
10^{-38}~{\rm s}.$

{\it ChT particles annihilations}

We have calculated that tensor particles annihilations proceed till
\begin{equation}
t_a\approx 2.42/(\sqrt{g_*}\:T_a^2{\rm [MeV]})~{\rm s}
\end{equation}
where  $T_a=2M_T$. In case  $M_T=3$ TeV is assumed $t_a\approx
5\times 10^{-14}~{\rm s}$. I.e. annihilation processes stop earlier
than estimated in previous works, which considered smaller ChT
particles masses.

{\it ChT particles decays}

 In the hot Universe plasma while relativistic the ChT particles do not
 effectively decay because their states are
repopulated by the inverse decay by the particles of the hot plasma,
so their density for $T > M_T$ is close to their equilibrium value.
With the decrease of the Universe temperature during its expansion
the ChT particles become non relativistic. Then the inverse decay
processes can be neglected. The cosmic time corresponding to the
decay of ChT particles at $T \sim M_T$ is $t_d\approx
2.42/(\sqrt{g_*}\:T_d^2{\rm [MeV]})~{\rm s}$, where $T_d\sim M_T$.
For  $M_T=3$ TeV $t_d\approx 2\times 10^{-13}~{\rm s}$.

The decay width of the tensor particles at rest is estimated to be:
\begin{equation}
\Gamma\approx g_T^2 M_T/4 \pi \approx 102~{\rm GeV}.
\end{equation}
This width is considerably larger (an order of magnitude) than the
estimated is earlier works. The lifetime is also changed, namely,
$\tau=6.5\times 10^{-27}$~s

Having in mind that the decay time is later than the annihilation
time, it can be concluded that the main part of the ChT particles
disappear from the cosmic plasma due to their annihilations at
$t\sim t_a$ and the rest decay rapidly soon after that at $t \sim
t_d$. Hence, the period of their
 abundant presence in the hot Universe is the period from the time of their
 creation to the time of their annihilation:
$$6\times 10^{-42}~{\rm s} <t<5\times 10^{-14}~\rm{s}.$$
  The corresponding energy
range is from $1.8 \times 10^{17}$~GeV to $6\times 10^3$~GeV.

The ChT particles disappear at early epoch and, therefore, they
 cannot disturb Big Bang Nucleosynthesis and Cosmic Microwave Background formation
epoch. On the other hand ChT particles are present at energies
typical for inflation, Universe reheating, lepto- and baryogenesis.
The extended model with ChT particles proposes new source for
CP-violation and may present a natural mechanism for leptogenesis
and baryogenesis scenarios.

\section{Conclusion}

We discuss an extended Beyond Standard Model of Particle Physics and
Cosmology with new ChT particles. The influence of these particles
on the early Universe is studied.

 At present the search of these particles is
conducted by the ATLAS Collaboration at LHC. First experimental
results provided constraints on the tensor particle masses and
couplings. Using current experimental and theoretical findings we
calculated the ChT particles characteristic processes:
     creation, scattering, annihilation and decay.

The characteristic interactions of the chiral tensor particles in
the early Universe plasma were found to be noticeably different than
previously calculated. The time interval of abundant presence of ChT
particles in the Universe evolution is determined. It lasts from the
time of their creation till their annihilation, namely: $6\times
10^{-42}~{\rm s}<t<5\times 10^{-14}~{\rm s}$. The corresponding
energy range is from $1.8\times 10^{17}$~GeV down to $6 \times
10^3$~GeV, which according to us is very promising for theoretical
speculations involving ChT particles concerning inflationary models,
reheating scenarios, baryogenesis, leptogenesis scenarios, etc.

The speeding up of the Hubble expansion due to the density increase
caused by the introduction of the new particles and the change of
the temperature-time dependence are estimated.

The discussed model of BSM physics with additional chiral tensor
bosons is
     allowed from cosmological point of view and hopefully its unique predictions
     will be tested at the new run of LHC.

\section*{Acknowledgments}
We are grateful to the unknown referee for the useful comments and
valuable criticism, which considerably helped to improve the paper.
This work was supported by the project DN 08-1/2016 of the National
Science Fund of the Bulgarian Ministry of Education and Science.


\begin{thebibliography}{0}

\bibitem{MPLA} M. V. Chizhov, {\it Mod. Phys. Lett. A} {\bf 8}, 2753 (1993).
\bibitem{EChAYa} M. V. Chizhov, {\it Sov. J. Part. Nucl.}, {\bf 26}, 553 (1995).
\bibitem{EChAYaa} M. V. Chizhov, {\it Sov. J. Part. Nucl.}, {\bf 42}, 93 (2011).
\bibitem{LHC} M. V. Chizhov, V.A. Bednyakov and J.A. Budagov, arXiv:0801.4235 {\it Sov. J. Part. Nucl.}, {\bf 71}, 2096 (2008);
 \\ M. V. Chizhov, arXiv:0807.5087 [hep-ph].
\bibitem{mih2009} M. Chizhov, {\it Phys.Part.Nucl.Lett.}{\bf 6}, 361 (2009).
\bibitem{dvali} M. Chizhov, G. Dvali, {\it Phys. Lett.B}{\bf 703}, 593 (2011)
\bibitem{ATLAS14} Aad, G. et al.(ATLAS Collab.), {\it Phys.Rev.D}{\bf 90}, 05, 205 (2014).
\bibitem{ATLAS14a} Aad, G. et et al. (ATLAS Collab.), {\it JHEP}{\bf 09}, 037 (2014a).
\bibitem{todor} D. P. Kirilova, M. V. Chizhov and T. V.
Velchev, {\it Comptes Rendus de l'Acad\'emie bulgare des Sciences}
{\bf 48}, 25 (1995)
\bibitem{AATr} D. P. Kirilova and M. V. Chizhov, {\it
Astronomical and Astrophysical Transaction} {\bf 3}, 205 (1998).
\bibitem{blois}Kirilova, D., and M. Chizhov (2009), A Possible TeV Window on the
Universe , in Proceedings of XXIth Rencontres de Blois "Windows on
the Universe"21-26 June, 2009, in the Chateau de Blois, Loire
Valley, France, edited by L. Celnikier, J. Dumarchez, and J. Tran
Thanh Van, Moriond Astrophysics Meeting, The Gioi Publishers,
Vietnam GPXB 4 - 1000/XB-QLXB.
\bibitem{IJMPA} M.V. Chizhov and D. Kirilova, {\it Int. J. Mod. Phys.} A {\bf 24}, 1643 (2009), arXiv:0903.4180.
\bibitem{dol} A. Dolgov, {\it Phys. Rept.} 370, 333 (2002).
\bibitem{2higgs} S. L. Glashow and S. Weinberg, {\it Phys. Rev. D}
{\bf 15}, 1958 (1977).
\bibitem{Avdeev} M. Chizhov, L. Avdeev, {\it Phys. Part. Nucl. Lett.}
2, (2005).
\bibitem{NJL} M. V. Chizhov, {\it JETP Lett.} {\bf 80}, 73 (2004).
\end{thebibliography}
\end{document}